\documentclass{PoS}

\usepackage[utf8]{inputenc}
\usepackage{amsmath,amsfonts,amssymb}

\usepackage{tikz}
\usetikzlibrary{patterns}
\title{Dynamics of branes in Double Field Theory}

\ShortTitle{Dynamics of branes in DFT}

\author{\speaker{Edvard T. Musaev}\\
        Moscow Institute of Physics and Technology\\Institutskii per. 9, Dolgoprudny, 141700, Russia\\
        E-mail: \email{musaev.et@phystech.edu}}

\abstract{In this contribution the construction of the Wess-Zumino term for Dp-branes and NS 5-branes invariant under T-duality is described. The cases of the massless Type IIA/B and the massive Type IIA theories are covered. Mass parameter is introduced by generalized Scherk-Schwarz ansatz of O(10,10) spinor and tensor fields. The results lead to the picture where a given brane is understood as an invariant object rotated into the full doubled space.}

\FullConference{Corfu Summer Institute 2018 "School and Workshops on Elementary Particle Physics and Gravity"\\
		(CORFU2018)\\
		31 August - 28 September, 2018\\
		Corfu, Greece}

\def\a{\alpha} \def\b{\beta} \def\d{\delta} \def\e{\epsilon}\def\h{\eta}
\def\k{\kappa} \def\l{\lambda}  \def\x{\xi}    \def\f{\varphi}
\def\ff{\phi}  \def\y{\psi} 

\def\G{\Gamma}\def\X{\Xi}

\def\C{{\mathcal{C}}}
\def\mG{\mathcal{G}}
\def\mH{\mathcal{H}}
\def\mF{\mathcal{F}}

\def\tx{\tilde{x}}
\def\tz{\tilde{z}}
\def\tdt{\tilde{\partial}}

\def\XX{\mathbb{X}}

\def\ddt{\partial\mkern-10mu/} 
\def\fr{\frac} \def\dt{\partial}

\begin{document}

\section{Introduction}

Among solutions of supergravity equations of motion one finds brane backgrounds, which preserve half of the total amount of supersymmetry. These are described by non-trivial metric and gauge fields sourced by the corresponding fundamental objects of string  or M theory \cite{Polchinski:1995mt}.  Despite the name, string theory contains information about dynamics of all such extended objects, which include D-branes, NS5-brane and the KK-monopole. These are usually dubbed the standard branes as opposed to the so-called non-standard or exotic branes that are related to the standard ones by T- and S-duality transformations\footnote{Strictly speaking KK-monopole should also be referred to as a non-standard brane as it has a special direction realized by the Hopf cycle.} \cite{Bergshoeff:2011zk,Bergshoeff:2011ee,deBoer:2012ma}. 

T-duality is a symmetry of the closed string on a torodial background rotating string winding and momentum modes into each other. For a torus $\mathbb{T}^d$ the rotation group is O$(d,d)$, which contains global coordinate transformations and constant B-shifts in addition to actual T-duality transformations. At the level of supergravity T-duality is a solution generating transformation and is allowed to act along isometries of a background. The corresponding transformation of the background is encoded in the so-called Buscher rules \cite{Buscher:1987qj,Duff:1989tf}. A relevant example would be the T-duality transformation of the background of smeared NS5-brane along the transverse isometry direction, which gives the KK5-monopole background. The isometry direction of the NS5-brane becomes the special cycle of the monopole (Hopf fibre). 
\begin{table}[http]
\centering
\begin{tabular}{|r|cccccc|cccc|}
\hline
           &       0&  1 &        2 &        3 &        4 &        5 &       6 &      7 &        8 & 9         \\  
 \hline
NS5=$5_2^0$   & $\times$    & $\times$ & $\times$ & $\times$ & $\times$ & $\times$     & $\cdot$ &$\cdot$ & $\cdot$  & $\cdot$  \\
KK5=$5_2^1$        & $\times$    & $\times$ & $\times$ & $\times$ & $\times$ & $\times$     & $\odot$ &$\cdot$ & $\cdot$  & $\cdot$  \\
 $5^2_2$   & $\times$    & $\times$ & $\times$ & $\times$ & $\times$ & $\times$     & $\odot$ &$\odot$ & $\cdot$  & $\cdot$  \\
  $5^3_2$   & $\times$    & $\times$ & $\times$ & $\times$ & $\times$ & $\times$     & $\odot$ &$\odot$ & $\odot$  & $\cdot$  \\
   $5^4_2$   & $\times$    & $\times$ & $\times$ & $\times$ & $\times$ & $\times$      & $\odot$ &$\odot$ & $\odot$  & $\odot$  \\
 \hline 
\end{tabular}
\caption{\sl T-duality orbit of the NS5-brane. Here $\times$ denote worldvolume directions of the brane, dots $\cdot$ denote transverse directions, which enter into dependence of the harmonic function of the corresponding background. Doted circle $\odot$ denotes special directions, which encode non-trivial monodromy around a brane. } 
\end{table}
Smearing the KK-monopole along a transverse direction and further applying T-duality transformation one arrives at a configuration of metric and Kalb-Ramond field which is not globally well defined. That is, the  background has non-trivial monodromy when going around the brane positioned at the singularity, and the solution is defined only up to a T-duality transformation and is described by T-folds \cite{Shelton:2005cf,Hull:2006qs}. Such non-geometric backgrounds are understood as sourced by exotic (or non-standard) branes, in this case, the $5_2^2$-brane \cite{deBoer:2012ma}. The notation $b_\alpha{}^c$ comes from the classification of states of half-maximal $D=3$ supergravity, which correspond to completely wrapped branes of co-dimension $\geq 2$. Mass of a state $b_\alpha{}^c$ will be
\begin{equation}
M_{b_\alpha{}^c }=\fr{R_1\dots R_b R_{b+1}^2\dots R_{b+c}^2}{g_s^{\alpha}l_s^{1+b+2c}},
\end{equation}
where $g_s$ is the string coupling constant, $l_s$ represents the Planck mass and $R_i$ are radii of compact directions. In other words, $b$ gives the number of radii entering linearly and $c$ gives the number of radii entering quadratically. In the language of extended objects these are the radii wrapped by worldvolume and special circles of the corresponding brane. Hence, the $5_2^2$-brane has 5+1 worldvolume directions, 2 special (quadratic) directions and its tension is proportional to $g_s{}^{-2}$. In these notations the NS5-brane and KK5-monopole are denoted as $5_2^0$ and $5_2^1$ respectively.

As dynamical objects branes interact with gauge potentials appearing as magnetic duals of the graviton and Kalb-Ramond field and/or their S/T-duals. The NS5-brane is known to interact with the 6-form potential $B_{(6)}$ which is the magnetic dual of the Kalb-Ramon 2-form field (see Section \ref{sec:WZstandard} for more details). Hence the brane interacts electrically with the 7-form field strength $H_{(7)}$ and magnetically with the usual 3-form field strength $H_{(3)}$. Smeared NS5-brane sources the background of the magnetic monopole which has only components $H_{mnz}$ non-zero, where $z$ is the smearing direction. T-dualizing along $z$ one arrives at the KK5-monopole, fro which the Kalb-Ramond field vanishes $B=0$ and the background is described completely in terms of space-time metric as a 9+1-dimensional Taub-NUT solution \cite{Gross:1983hb}. The gauge field flux $H_{mnz}$ gets transformed into the geometric flux $f_{mn}{}^z$. At the linearized level the latter can be understood as a proper field strength given by a space-time 2-form taking values in 1-vectors. Following the same Hodge dualization procedure as for the Kalb-Ramond field one concludes, that the corresponding gauge potential is given by a (7,1) mixed symmetry tensor $A_{m_1\dots m_7,n}$ \cite{Kleinschmidt:2011vu}. The additional condition here is that the only non-vanishing components are those for which $n$ equals to one of $m_i$'s. One writes then for the field strength $f_{mn}{}^z=g^{zp}F_{mn,p}$
\begin{equation}
F_{mn,p}=\e_{mn}{}^{m_1\dots m_8}\dt_{m_1}A_{m_2\dots m_8,p}.
\end{equation}
Note that we are working here at the linearized level. Similarly, Q-flux $Q_m{}^{pq}$ and R-flux $R^{mnp}$ can be understood as field strengths for the (8,2) and (9,3) gauge potentials respectively. These interact with the $5_2^2$ and $5_2^3$ branes. The case of co-dimension-0 brane $5_2^4$ is subtle as one does not have enough coordinates do define field strength as derivative of a (10,4) potential, and should be addressed in the framework of doubled space.

In the approach of conventional supergravity T-duality is a symmetry of toroidally compactified theory and is allowed to act only along isometric directions. This is relaxed in the T-covariant approach of Double Field Theory, where T-duality transformations appear as a subset of local coordinate transformation of the doubled space  \cite{Hohm:2010pp,Hohm:2010jy} (for a review see \cite{Berman:2013eva,Aldazabal:2013sca}). The doubled space is parametrized by coordinates $\XX^M=(x^m,\tx_m)$, with $x^m$ being the normal coordinates and $\tx_m$ being the dual coordinates corresponding to winding modes of the string. In the full O$(10,10)$ DFT the metric and Kalb-Ramond field are combined into an element of the coset space $\rm{O}(10,10)/\rm{O}(1,9)\times \rm{O}(9,1)$ called generalized metric $\mH_{MN}$ and a T-duality scalar $d=\f-1/4 \log \det g$ called the invariant dilaton. The defining feature of the theory is that all fields are allowed to depend on all coordinates $\XX^M$, however subject to a special constraint, the section condition, which can be schematically written as
\begin{equation}
\h^{MN}\dt_M\otimes \dt_N=0.
\end{equation}
Here $\h^{MN}$ is the invariant tensor of O$(10,10)$ which is normally taken in the light-cone basis
\begin{equation}
\h^{MN}=
\begin{bmatrix}
\bf{0} & \bf{1} \\
\bf{1} & \bf{0}
\end{bmatrix}.
\end{equation}
This tensor is also used to raise and lower indices. The section constraint is necessary to ensure closure of algebra of generalized Lie derivatives defining an analogue of diffeomorphisms in doubled space \cite{Berman:2012vc}. A finite version of coordinate transformation in doubled space is also available \cite{Hohm:2012gk,Rey:2015mba}.

Since fields of DFT are allowed to depend maximally on the full set of ten  (normal) coordinates, nature of T-duality symmetry is very different for the conventional supergravity and DFT. T-duality transformation in Double Field Theory in a direction $x$ can be performed even if the direction is non-isometric, and is equivalent to replacing $x$ by its dual $\tx$. Apparently, such transformation is solution generating in DFT, however it gives solutions of normal supergravity only if $x$ is an isometric direction. For toroidal backgrounds both DFT and supergravity approaches give the same results.

An illustrative example of this feature is provided by the T-duality orbit starting with the NS5-brane, which is lifted into DFT as the so-called DFT monopole \cite{Berman:2014jsa}. Considering a formal expression $ds^2=\mH_{MN} d\XX^M d\XX^N$ and requiring its invariance under T-duality of DFT, one is able to recover NS5-brane, KK5-monopole and the exotic branes $5_2^2$, $5_2^3$, $5_2^4$ by properly replacing geometric coordinates by their duals \cite{Jensen:2011jna,Bakhmatov:2016kfn}. The important issue here is that the resulting backgrounds do not have isometries along the would be special directions, instead, these are represented by dependence on the corresponding dual coordinates. For example, the localized KK5-monopole background is obtained from the full unsmeared NS5-brane background by replacing a transverse coordinate $z$ by its dual $\tz$. The resulting background is still a solution of DFT, however no longer solves supergravity equations of motion. Nevertheless, the result still makes sense as precisely the same background has been shown to arise after taking into account instanton correction to the usual KK5-monopole background \cite{Tong:2002rq,Harvey:2005ab,Jensen:2011jna}. The same is true for other exotic branes of the orbit, whose background fields will depend on 2,3 and 4 dual coordinates. This result has also been checked by a direct calculation of instanton corrections \cite{Kimura:2013zva,Kimura:2013fda}.

The same approach has been developed for the Exceptional Field Theory, which is the M-theory counterpart of DFT \cite{Hohm:2013pua,Hohm:2014fxa,Hohm:2013uia,Hohm:2013vpa,Abzalov:2015ega,Musaev:2015ces,Hohm:2015xna}. In this approach exotic branes of M-theory appear as a single object rotated in the full extended space with fields allowed to depend on non-geometric coordinates \cite{Bakhmatov:2017les,Berman:2018okd,Fernandez-Melgarejo:2018yxq} (for a short review of the classification of branes in the $E_7$ theory see \cite{Otsuki:2019owg}).

In the covariant approach of DFT the background of the DFT monopole is sourced by an extended  object whose worldvolume action drops into that of the NS5-brane and KK5-monopolke depending on the choice of embedding into the full doubled space. For the DBI action this has been shown in \cite{Blair:2017hhy} by introducing generalized Killing vectors that define charge of the object for a given embedding and, more important, select isometric directions. Existence of isometries is a direct consequence of the section constraint, which demands that one must choose whether a background has isometry in the normal $x$ or the dual $\tx$ space by defining the charge or, equivalently, the generalized Killing vector. To couple such brane action to the field theory action of DFT one must decide which coordinates in the DFT action are geometric and which are dual, then various choices of the generalized Killing vectors will produce the corresponding localized (exotic) brane backgrounds with proper dependence on dual coordinates.

In this contribution we briefly review the results of the work \cite{Bergshoeff:2019sfy} where T-duality covariant expressions for the Wess-Zumino terms for Dp-branes, NS 5-branes and $\a=3,4$ branes have been presented. We start with details of the conventional supergravity approach to construction of gauge invariant Wess-Zumino terms for D-branes and the NS5-brane in Section \ref{sec:WZstandard}. Section \ref{WZ} describes construction of the T-duality invariant expression Wess-Zumino terms for $\a=1$ (D-branes) and $\a=2$ branes. In this section embedding of a brane into the full doubled space is encoded in a gauge fixing condition. Finally, in Section \ref{concl} we summarize the results, discuss further developments and speculate around the topic.

\section{Wess-Zumino terms and supergravity}
\label{sec:WZstandard}

The action for the Type IIB supergravity in the string frame can be written as
\begin{equation}
\begin{aligned}
S_{RR}^{IIB}= \fr{1}{2\k_{10}^2}\int_{M_{10}}&\bigg[e^{-2\ff}\Big(*1 R-4d\ff\wedge *d\ff+\fr12 H_3 \wedge*H_3\Big)\\
&-\fr12G_2\wedge *G_2-\fr12G_4\wedge * G_4+\fr12dC_3\wedge dC_3 \wedge B_2\bigg],
\end{aligned}
\end{equation}
where $\k_{10}$ is the gravitational coupling in ten dimensions, the gauge invariant field strengths $G_{2,4}$ are defined as
\begin{equation}
\begin{aligned}
G_2&=dC_2,\\
G_4&=dC_3+H_3\wedge C_1
\end{aligned}
\end{equation}
and the NS-NS field strength $H_3=dB_2$. To end up with the democratic formulation of the theory where the fields $C_1,C_3$ and their magnetic duals enter on equal footing, one starts with equations of motion and Bianchi identities and interprets both as equations of motion following from a pseudo-action. Hence, we write for the EoMs for the fields $C_1,C_3$
\begin{equation}
\begin{aligned}
d*G_2-H_3\wedge *G_4&=0,\\
d*G_4-dC_3\wedge H_3&=0.
\end{aligned}
\end{equation}
From the second line one concludes $d(*G_4+H_3\wedge C_3)=0$ and defining $G_6=-*G_4$ introduces gauge potential $C_5$ as
\begin{equation}
G_6=dC_5+H_3\wedge C_3.
\end{equation}
The same procedure for the first line gives definition of the gauge potential $C_7$ and the corresponding field strength
\begin{equation}
G_8=dC_7+H_3\wedge C_5.
\end{equation}
Altogether the gauge invariant RR field strengths and the corresponding Hodge dualizations can be summarized as follows
\begin{equation}
\begin{aligned}
G_p&=dC_{p-1}+H_3 \wedge C_{p-3},\\
*G_p&=(-1)^{\fr p2+1} G_{10-p}.
\end{aligned}
\end{equation}
In the democratic formulation the Type IIB supergravity action for the RR fields is replaced by the following pseudo-action
\begin{equation}
S_{RR}^{pA}=\int_{M_{10}}\Big[-\fr12G_2\wedge *G_2-\fr12G_4\wedge * G_4-\fr12G_6\wedge *G_6-\fr12G_8\wedge * G_8\Big],
\end{equation}
whose equations of motion give equations of motion for the fields $C_1,C_3$ and the corresponding Bianchi identities upon the duality relations $*G_p=(-1)^{\fr p2+1} G_{10-p}$. Note that according to the pseudo-action formalism  the duality relation must be imposed only at the level of equations of motion, since the action simply vanishes on this relation. 

The field strengths $G_p$ are invariant under gauge transformations of the RR fields
\begin{equation}
\d C_p= d\l_{p-1}+H_3\wedge \l_{p-3}.
\end{equation}
This leads to subtleties in definition of charge for a D-brane \cite{Marolf:2000cb} and in turn to the Hanany-Witten effect \cite{Hanany:1996ie}. Such non-trivial gauge transformation of the RR field does not allow to write Wess-Zumino term for a Dp-brane naively as 
\begin{equation}
S^{(0)}_{WZ}=\int d^{p+1}\x C_{p+1},
\end{equation}
since it will not be gauge invariant. Indeed, under gauge transformation of $C_p$ the action transforms as
\begin{equation}
\d S_{WZ}^{(0)}= \int d^{p+1}\x\,  H_3 \wedge \l_{p-2},
\end{equation}
which suggests to add a term of the form $B_2 \wedge C_{p-1}$ to the initial action and to consider instead
\begin{equation}
S^{(1)}_{WZ}=\int d^{p+1}\x \Big(C_{p+1} +B_2 \wedge C_{p-1}\Big).
\end{equation}
Pieces coming from transformation of the second term cancel the terms appeared at the previous step, however generating new unwanted terms. Following the same logic it is straightforward to check that the Wess-Zumino action invariant under gauge transformations of the RR fields should be of the form
\begin{equation}
\tilde{S}_{WZ}=\int d^{p+1}\x \Big[e^{B_2}\wedge \C\Big]_{p+1-\,\rm{form}}.
\end{equation}
Here we define the multiform $\C$ as the following formal sum of all RR fields of the theory
\begin{equation}
\C=C_1+C_3+C_5+C_7,
\end{equation}
and the resulting expression should be projected to the space of $p+1$ forms. Finally, one notices that the action $\tilde{S}_{WZ}$ is not invariant under gauge transformations of the Kalb-Ramond field $\d B_{2}= d\L_1$. This is fixed along the same line as for the open string action, i.e. replacing $B_2 \to \mF_2=d A_1+B_2$, where $A_1=A_\a d\x^\a$ is the worldvolume 1-form interacting with open strings ending on the D-brane transforming as
\begin{equation}
\d A_1=-\L_1.
\end{equation} Here $B_2$ is understood as the worldvolume pull-back $B_{\a\b}d\x^\a \wedge d\x^\b$ of the target space 2-form. Hence, the Wess-Zumino action for Dp-brane invariant under gauge transformations of both the RR fields and the Kalb-Ramond field can be rendered as
\begin{equation}
\label{WZDp}
S_{WZ}^{Dp}=\int d^{p+1}\x \Big[e^{\mF_2}\wedge \C\Big]_{p+1-\,\rm form}.
\end{equation}

Similar procedure can be performed for  $B_2$ to recover its magnetic dual interacting with the NS5-brane. One starts with equations of motion for the 2-form NS-NS field
\begin{equation}
d(e^{-2\ff }*H_3)+d\Big(C_1\wedge G_5-C_3\wedge G_4+C_5\wedge G_5\Big)=0.
\end{equation}
Hence, the magnetic 7-form NS-NS flux is defined as $H_7=e^{-2\ff}*H_3$ and the corresponding 6-form gauge field $B_6$ is defined by the field strength as
\begin{equation}
\label{H7}
H_7=dB_6-C_1\wedge G_5+C_3\wedge G_4-C_5\wedge G_5,
\end{equation}
which is trivially invariant under gauge transformations $\d B_6=d\L_5$. To ensure invariance under gauge transformations of the RR fields one starts with the full gauge transformations of the magnetic 6-form defined as
\begin{equation}
\d B_6=d\L_5+\l_0\wedge G_6-\l_2\wedge G_4+\l_4\wedge G_2.
\end{equation}
As in the case of D-branes the naive Wess-Zumino term for the NS5-brane $\tilde{S}_{WZ}^{NS5}=\int d^6\x B_6$ is not gauge invariant under $\l_p$, instead its transformation reads
\begin{equation}
\d\tilde{S}_{WZ}^{NS5}=\int d^6\x\Big(\l_0\wedge G_6-\l_2\wedge G_4+\l_4\wedge G_2\Big).
\end{equation}
To fix this one needs to introduce extra worldvolume fields $c_p$ whose transformation rules are defined as
\begin{equation}
\begin{aligned}
& \d c_0=-\l_0, && \d c_2=-\l_2, && \d c_4=-\l_4. && 
\end{aligned}
\end{equation}
The fields $c_p$ interact with Dp-branes ending on the NS5-brane. Finally, in analogy with the worldvolume field strength $\mF=dA+B_2$ of the NS-NS sector one defines field strengths
\begin{equation}
\mG_p=dc_p+C_p+H_3\wedge c_{p-3},
\end{equation}
which are invariant under gauge transformations of the RR fields. Altogether this allows to define gauge invariant Wess-Zumino term for the NS5-brane
\begin{equation}
\label{WZNS5}
S_{WZ}^{NS5}=\int d^6\x \Big(B_6-\mG_1\wedge C_5+\mG_3\wedge C_3-\mG_5\wedge C_1\Big).
\end{equation}

Smearing the background of the NS5-brane generates an isometry direction along which a T-duality transformation can be preformed. This leads to the background of the KK5-monopole characterized by the geometric flux $f_{mn}{}^z$, where $z$ is the isometry direction and $m,n\neq z$. Following the same dualization procedure as the one for the Kalb-Ramond field strength one arrives (at the linearized level) at the gauge potential $B_{m_1\dots m_6,n}$, where $n$ must be equal to one of the $m$'s. This is a (7,1) mixed symmetry gauge field interacting with the KK5 monopole \cite{Kleinschmidt:2011vu}. 

The same procedure can be done for the Wess-Zumino action of the NS5-brane to generate that of the KK5-monopole \cite{Eyras:1998kx}. Following the T-duality orbit further one generates Wess-Zumino terms for the exotic branes $5_2^2$ and $5_2^3$ \cite{Chatzistavrakidis:2013jqa}. These interact with gauge potentials $B_{8,2}$ and $B_{9,3}$, hence showing two and three special directions respectively. In O(d,d)-covariant theory these transform as a rank-4 antisymmetric tensor multiplet and can be collected into components of a tensor $D_{MNPQ}$ as
\begin{equation}
\begin{aligned}
5_2^0:&& D^{mnpq}&=\e^{mnpqm_1\dots m_6}B_{m_1\dots m_6},\\
5_2^1:&&D^{mnp}{}_{q}&=\e^{mnpm_1\dots m_7}B_{m_1\dots m_7,q},\\
5_2^2:&&D^{mn}{}_{pq}&=\e^{mnm_1\dots m_8}B_{m_1\dots m_8,pq},\\
5_2^3:&&D^{m}{}_{npq}&=\e^{mm_1\dots m_9}B_{m_1\dots m_9,npq},\\
5_2^4:&&D_{mnpq}&=\e^{m_1\dots m_{10}}B_{m_1\dots m_{10},mnpq}.
\end{aligned}
\end{equation}
Here we also included the co-dimension-0 brane $5_2^4$ for completeness. In the next Section we describe the construction of the $O(d,d)$ invariant expressions for D-branes and the NS5-brane orbit, reviewing the work \cite{Bergshoeff:2019sfy}.

\section{Invariant Wess-Zumino actions}
\label{WZ}

\subsection{D-branes}

Upon T-duality transformations the background of a Dp-brane is mapped into the background of D(p$\pm$1)-brane depending on the direction along which the transformation is performed. When T-dualizing along a compactified transverse direction one produces a brane of one dimension larger, and when dualizing along a wrapped worldvolume direction one produces a brane of one dimension lower. This can be summarized in the following T-duality rules for the RR potentials
\begin{equation}
\begin{aligned}
C_{m_1\dots m_p z} && \overset{T_z}{\longleftrightarrow} && C_{m_1\dots m_p}.
\end{aligned}
\end{equation}
This implies that at the DFT language RR potentials can be written as an O(10,10) spinor \cite{Hohm:2011dv}. For that one introduces algebra of O(10,10) gamma matrices 
\begin{equation}
\{\G_M,\G_N\}=2\h_{MN}.
\end{equation}
In the usual GL(10) decomposition this can be written as
\begin{equation}
\begin{aligned}
\{\G_m,\G^n\}=2\d_m{}^n.
\end{aligned}
\end{equation}
Introducing raising and lowering operators $\sqrt{2}\y^M=\G^M$ one defines Clifford vacuum $|0\rangle$as
\begin{equation}
\y_m |0\rangle=0.
\end{equation}
Hence, an O(10,10) spinor encoding all degrees of freedom coming from the RR gauge fields $C_{m_1\dots m_p}$ is defined as
\begin{equation}
C=\sum_{p=0}^{10}\fr1{p!}C_{m_1\dots m_p}\y^{m_1}\dots \y^{m_p}|0\rangle.
\end{equation}
The sum is performed along forms of all ranks and hence the spinor $C$ describes both IIA and IIB theory. Upon T-duality it transforms as
\begin{equation}
\begin{aligned}
T_z:&& C \to \G_z\, C.
\end{aligned}
\end{equation}
Chirality of the spinor C and of the vacuum $|0\rangle$ can be fixed independently, and it is convenient to fix chirality of $C$ to be positive and choose chirality of $|0\rangle$ depending on whether we are in the IIA or IIB theory. Apparently, such defined action of a T-duality transformation flips chirality of the vacuum.

To construct an O(10,10) covariant expression for the gauge invariant field strengths $G_{p}$ in this language we will need to define the following Clifford element
\begin{equation}
\begin{aligned}
S_B=e^{-\fr12 B_{m n}\G^m \G^n}&& \Longrightarrow && S_B \ddt S_B{}^{-1}=\fr16 H_{mnp}\G^{mnp},
\end{aligned}
\end{equation}
where $\ddt=\G^M\dt_M$. In what follows we will always impose the solution of the section constraint $\tdt^m=0$, and hence do not consider branes localized in the dual space\footnote{More comment on this in Section \ref{concl}}. Encoding the gauge transformation parameters $\l_{m_1\dots m_p}$ in a DFT spinor $\l$ in the same fashion, one writes gauge transformation of $C$ as
\begin{equation}
\d C=\ddt \l+S_B\ddt S_B{}^{-1} \l.
\end{equation} 
The gauge invariant field strength is then
\begin{equation}
G=\ddt C+S_B\ddt S_B{}^{-1} C,
\end{equation}
which is a DFT spinor of chirality opposite to that of $C$.

To encode the worldvolume gauge field $A_1$ in this language it is convenient to start with the case of the D9-brane, which is space-filling and hence the worldvolume coordinates can be gauge-fixed to coincide with the space-time coordinates $x^m$. In what follows we will always assume such gauge fixing for the D-branes. This allows to introduce the gauge invariant Clifford algebra element $S_{\mF}$ in analogy with $S_B$ as
\begin{equation}
S_{\mF}=e^{-\fr12 \mF_{mn}\G^m\G^n}.
\end{equation}
Hence, the Wess-Zumino term for the D9-brane in these variable can be written as follow
\begin{equation}
\label{WZ9}
S_{WZ}^{D9}=\int d^{10}\x \bar{Q}_{10}S_{\mF}{}^{-1}C,
\end{equation}
where the charge spinor $\bar{Q}_{10}$ is defined as
\begin{equation}
\bar{Q}_{10}=\fr{q}{2^{10}}\langle 0|\G_0\dots \G_9.
\end{equation}
The expression \eqref{WZ9} precisely reproduces \eqref{WZDp} for $p=9$.

Wess-Zumino term \eqref{WZ9} is written in DFT covariant terms and hence is expected to provide Wess-Zumino terms for all other Dp-branes upon T-duality rotations. The only issue here would be the gauge fixing condition that relates worldvolume and space-time coordinates.

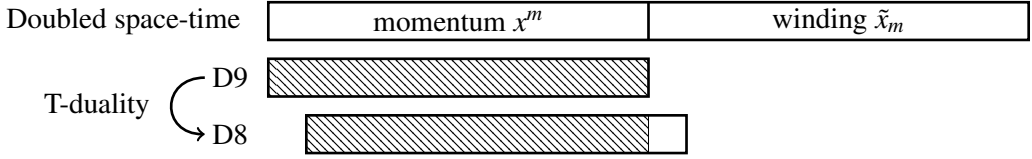
\begin{figure}
	\centering
	\begin{tikzpicture}
	\draw[line width=1pt] (0,2.75) -- (10,2.75) -- (10,3.25) -- (0,3.25)--cycle;
	\draw[line width=1pt] (5,2.75) -- (5,3.25);
	\node[align=center] at (2.5,3) {momentum $x^m$};
	\node[align=center] at (7.5,3) {winding $\tilde{x}_m$};
	\node[align=center] at (-1.9,3) {Doubled space-time};
	\draw[pattern=north west lines,line width=1pt] (0,2) -- (5,2) -- (5,2.5) -- (0,2.5)--cycle;
	\node[align=center] at (-0.5,2.25) {D9};
	\node (1) at (-0.7,2.25) {};
	\draw[line width=1pt] (0.5,1.25) -- (5.5,1.25) -- (5.5,1.75) -- (0.5,1.75)--cycle;
	\draw[pattern=north west lines,line width=0.1pt] (0.5,1.25) -- (5,1.25) -- (5,1.75) -- (0.5,1.75)--cycle;
	\node[align=center] at (-0.5,1.5) {D8};
	\node (2) at (-0.7,1.5) {};
	\draw[->,line width=1pt] (1) to [out=-180,in=180,looseness=2] (2);
	\node[align=center] at (-2.25,1.875) {T-duality};
	\end{tikzpicture}
	\caption{\label{fig:Tdual}\textit{D-branes in doubled space. All branes have a ten-dimensional worldvolume and the intersection of this with the ten physical momentum dimensions gives the apparent dimensionality of the worldvolume. T-duality along an isometry direction can move part of the ten-dimensional worldvolume between momentum and winding directions.}}
\end{figure}
Following the ideas expressed in \cite{Asakawa:2012px,Albertsson:2011ux} it is convenient to understand any Dp-brane as a D9 brane rotated in the full doubled space as depicted on Fig. \ref{fig:Tdual}. Hence, keeping the gauge fixing condition unchanged, one of the worldvolume coordinates, say $\x^9$, will be identified with a dual coordinate, say $\tx_9$. Since we always assume $\tdt^m=0$, this implies that no fields depend on $\x^9$ and the integral becomes trivial and can be dropped leaving us with one worldvolume integration less. This can be interpreted as changing the dimensionality of the worldvolume of the rotated D9 brane in the visible (geometric) space-time, which becomes worldvolume of a D8-brane.

Rotation of the D9 brane in the doubled space can be encoded in action of T-duality transformation on the charge $Q_{10}$
\begin{equation}
\bar{Q}_9= \bar{Q}_{10} \G^9=\fr{q}{2^9}\langle 0|\G_0\dots \G_8,
\end{equation}
where we act along $x^9$ for concreteness. It is straightforward to check that such transformed charge gives the correct Wess-Zumino action for the D8-brane
\begin{equation}
S_{WZ}^{D8}=\int d^{10}\x \bar{Q}_{9}S_{\mF}{}^{-1}C=\int d^9 \x \Big[e^{\mF_2}C\Big]_{\rm 9-form}.
\end{equation}
Here the integral over $\x^9$ is dropped since it is now gauge fixed to be $\tx_9$ and nothing depends on it. Also we assume proper normalization of the overall expression.

Hence, the final answer for the expression providing Wess-Zumino term for any Dp-brane is
\begin{equation}
\label{WZDinv}
S_{WZ}^{Dp}=\int d^{10}\x \bar{Q}_{p}S_{\mF}{}^{-1}C.
\end{equation}
This is T-covariant as is written in DFT notations and hence does not change its form upon T-duality transformations. To reproduce the Wess-Zumino term for a given Dp-brane one must choose the charge $Q_p$ appropriately. This can be interpreted as choosing a frame in which the D9-brane is rotated such that to be partially invaded into the doubled space.

Following the same scheme one introduces Wess-Zumino term for Dp-branes of massive Type IIA theory by a mild deformation of the RR spinor inside DFT
\begin{equation}
\label{SSC}
\begin{aligned}
C && \to && C+\fr{m}{2}S_B \tx_1 \G^1 |0\rangle,
\end{aligned}
\end{equation}
where $\tx_1$ is a coordinate in the dual space. This is the generalized Scherk-Schwarz mechanism suggested in \cite{Hohm:2011cp}. In principle one is allowed to take any dual coordinate instead of $\tx_1$, this will not change the final result. Expression for the Wess-Zumino term in the massive theory remains in the same covariant form \eqref{WZDinv} with $C$ replaced according to the Scherk-Schwarz ansatz above.

\subsection{NS 5-branes }

Gauge potentials interacting with branes of the T-duality orbit of NS5-brane can be encoded (at the linearized level) in the rank four antisymmetric tensor $D^{MNKL}$ of DFT. In addition to define dual formulation of Double Field Theory on introduces auxiliary potentials $D^{MN}$ and $D$, whose field strengths are written as \cite{Bergshoeff:2016ncb}
\begin{equation}
\begin{aligned}
H^{(0)}_{MNP}&=\dt^QD_{QMNP}+3\dt_{[M}D_{NP]},\\
H^{(0)}_{M}&=\dt^ND_{MN}+\dt_M D.
\end{aligned}
\end{equation}
The first line encodes the 3-form field strength $H_{mnp}$, geometric flux $f_{mn}{}^k$ as well as the exotic fluxes $Q_m{}^{kl}$ and $R^{mnk}$. 

These field strengths are invariant under gauge transformations
\begin{equation}
\begin{aligned}
 \delta D^{MNPQ}& =\partial_R \Xi^{RMNPQ} + 4 \partial^{[M} \Xi^{NPQ]}\,, \\
 \delta D^{MN} &=\partial_P \Xi^{PMN} + 2 \partial^{[M} \Xi^{N]}\,, \\
 \delta D &= \partial_M \Xi^M \, . 
\end{aligned}
\end{equation}
The important check of whether the gauge invariant Wess-Zumino term for NS5-brane \eqref{WZNS5} can be generalized to other branes of the orbit at the full non-linear level, is the possibility of adding expressions non-linear in the RR fields to the covariant field strengths $H_{MNP}$ and $H_M$. Straightforward calculation shows that because of the section constraint of DFT this is impossible for any component except $H_{mnp}$. This is consistent with the fact that the dualization procedure works only at the linear level.

To reproduce the gauge invariant field strength \eqref{H7} one considers the following DFT covariant expression
\begin{equation}
H_{MNP}=\dt^QD_{QMNP}+\bar{G}\G_{MNP}C
\end{equation}
with gauge transformation of the potential $D_{MNKL}$ defined as
\begin{equation}
\label{gaugeD}
\d D_{MNPQ}=\dt^R \X_{RMNPQ}+\bar{G}\G_{MNPQ}\l.
\end{equation}
Using the Bianchi identity $\ddt G=-S_B \ddt S_B{}^{-1}G$ one shows that such defined field strength $H_{MNP}$ is invariant under gauge transformations of the field $D_{MNKL}$ up to terms where derivative has free index
\begin{equation}
\d H_{MNP}\propto \dt_{[M}(\bar{G}\G_{NP]}\l).
\end{equation}
Clearly, for the field strength $H^{mnp}$ of the NS5-brane gauge potential $B_6$ these terms vanish upon the solution $\tdt^m=0$ of the section constraint. While the other components remain not gauge invariant.

As it is explicitly checked in \cite{Bergshoeff:2019sfy} gauge transformation \eqref{gaugeD} correctly reproduce those of the 6-form potential interacting with the NS5-brane and the (7,1) mixed symmetry potential interacting with the KK5-monopole. The same is true for the remaining mixed symmetry potentials of the orbit. These results allow to compose a Wess-Zumino term for the NS 5-branes which reproduces the correct form \eqref{WZNS5} for the case of the NS5-brane and gives expected results at the linearized level for other branes of the orbit. Hence, we write
\begin{equation}
\label{WZNS}
S_{WZ}^{\a=2}=\int d^6 \x Q_{MNPQ}\Big(D^{MNPQ}+\bar{\mG}\G^{MNPQ}C\Big),
\end{equation}
where the capital indices transform in the vector representation of O(4,4). This corresponds to the split O(10,10) $\hookleftarrow$ O(4,4)$\times$O(6,6) induces by the chosen embedding of the worldvolume of the brane. Gamma matrices are decomposed accordingly
\begin{equation}
\begin{aligned}
\G_{\hat{M}}=(\G_A,\G_M\G^*),
\end{aligned}
\end{equation}
where $\G_A$ are the O(6,6) gamma matrices and $\G^*$ is the O(6,6) chirality matrix. Here hatted indices label the full 10+10-dimensional doubled space.

The charge $Q_{MNPQ}$ must be chosen manually to reproduce expression for the Wess-Zumino term of a given brane. One associates the components $Q_{mnpq}$ with the NS5-brane, $Q_{mnp}{}^q$ with the KK5-monopole and so on.

The worldvolume field strength $\mG$ is defined as
\begin{equation}
\mG=\ddt c+C+S_B\ddt S_B{}^{-1} c.
\end{equation}
Important obstruction to claim that \eqref{WZNS} provides full Wess-Zumino term for $\a=2$ branes is that the expression in parentheses is not gauge invariant. Indeed, one shows that the transformation
\begin{equation}
\d \Big(D^{MNPQ}+\bar{\mG}\G^{MNPQ}C\Big)
\end{equation}
vanishes only if all indices are up and if all down indices correspond to isometric directions. The interpretation is that the expression \eqref{WZNS} can be understood as a proper Wess-Zumino term for the T-duality orbit starting with the NS5-brane only if the background of the $5_2^r$-brane has $r$ isometric directions. This is in consistency with the common results in the literature \cite{Chatzistavrakidis:2013jqa}.

It is straightforward to write Wess-Zumino term for the NS 5-branes in the massive Type IIA theory following Scherk-Shwarz ansatz similar to \eqref{SSC}, which for the NS-NS magnetic gauge fields becomes
\begin{equation}
\begin{aligned}
\hat{D}^{MNPQ} (x,\tilde{x})&= D^{MNPQ} (x) + \frac{m}{2}  \tilde{x}_1 \langle 0| A\ \Gamma^1 \Gamma^{MNPQ} C \,, \\
\hat{D}^{MN} (x,\tilde{x}) &= \frac{m}{2} \tilde{x}_1 \langle 0| A\ \Gamma^1 \Gamma^{MN} C \, .\label{SSansatzDMNPQDMN}
\end{aligned}
\end{equation}
Expression for the field strength $H^{MNP}$ remains the same in terms if these new fields $\hat{D}^{MNPQ}$ and $\hat{D}^{MN}$, and in terms of the undeformed fields it becomes
\begin{equation}
H^{MNP} = \partial_Q D^{QMNP} + \overline{{C}} \Gamma^{MNP} {\partial\mkern-10mu/} {C} + 2 m \overline{C} \Gamma^{MNP} | 0\rangle \quad . \label{finalHromansappendix}
\end{equation}
As before the procedure does not spoil the components $H^{mnp}$, and these do not gain dependence on the dual coordinate $\tx_1$. To keep this dependence out of the remaining coordinates as well, one should always assume existence of $r$ isometries for component with $r$ lower indices.

The Scherk-Schwarz ansatz extended to the  worldvolume field strength $\mG$ becomes
\begin{equation}
\begin{aligned}
{\cal G} = &\ \ddt c + C + S_B \ddt S_B^{-1} c \\
&+ m b_A \Gamma^A \sum_{N=0}^3 \frac{1}{N+1} \left( S_B^{(N)} + \frac{1}{N} \sum_{n=1}^{N-1} S_B^{(N-n)} S_{\cal F}^{(n)} + S_{\cal F}^{(N)}  \right) | 0\rangle,
\end{aligned}
\end{equation}
where $S_B^{(N)}$ denotes rank $N$ form in expansion of 
the exponent.

Collecting these results one finds the following expression for the Wess-Zumino term
\begin{equation}
\label{WZNSm}
S_{WZ}^{{\rm NS5}m} = \int d^{6}\xi \  Q_{{M}{ N}{P}{Q} } \Big[  D^{{M}{N}{P}{Q}} +  \overline{\cal G} \Gamma^{{M}{N}{P}{Q}} C  -m \overline{c} \Gamma^{{M}{N}{P}{Q}} (S_B + S_{\cal F} )  | 0\rangle \Big]  . 
\end{equation}

\section{Conclusions}
\label{concl}

In this contribution the results of the work \cite{Bergshoeff:2019sfy} are briefly reviewed. We describe DFT covariant constructions for Wess-Zumino actions for branes with tension scaling as $g_{s}^{-\a}$ with $\a=1,2$. These correspond to D-branes and to NS 5-branes. We describe the constructions for the massless theory in \eqref{WZDinv} and \eqref{WZNS}, and for the massive Type IIA theory in \eqref{SSC} and \eqref{WZNSm}.

The obtained expression for the Wess-Zumino term for the NS 5-branes suggests interpretation of a given $5_2^r$-brane as a rotation of the (smeared) NS5-brane in the full doubled space. The presented procedure requires smearing when moving along the T-duality orbit since expressions for the field strengths entering the Wess-Zumino term are gauge invariant only if the $5_2^r$-brane background has $r$ isometries. The same condition prevents from constructing a full non-linear gauge invariant field strength $H^{mn}{}_p$ for the KK5-monopole, which is in agreement with the common understanding of the dual graviton problem\footnote{See \cite{Hohm:2018qhd} for more discussion on the current status of the issues with the dual graviton.}.

On the other hand the works \cite{Berman:2014jsa,Jensen:2011jna,Bakhmatov:2016kfn,Harvey:2005ab,Blair:2017hhy,Kimura:2016anf} for the branes of string theory and \cite{Bakhmatov:2017les,Berman:2018okd,Fernandez-Melgarejo:2018yxq} for M-theory  suggest that one could interpret the instanton-corrected background of say the KK5-monopole as the full NS5-brane background rotated in the doubled space without any isometries assumed. It is an open question how one lifts this low energy picture to the full brane worldvolume actions keeping gauge invariance and not interfering with the dual graviton problem.

The picture becomes even more intriguing when one turns to the  case of Dp-branes. In the construction above one starts with the Wess-Zumino term for the D9-brane and then rotates it in the full doubled space introducing isometries at each step. However, in analogy with the localized NS 5-brane solutions one may consider localized Dp-brane solutions, with bulk fields depending on the $9-p$ transverse geometric coordinates and $p$ dual coordinates (we do not count $\tx_0$ to avoid issues with timelike T-duality). Although the corresponding Wess-Zumino term straightforwardly follows from the expression \eqref{WZDinv}, it is not completely clear how to write the DBI action. The situation here is the opposite of that of the NS 5-branes, where the invariant DBI action producing localized backgrounds has been presented in \cite{Blair:2017hhy}, while to write gauge invariant Wess-Zumino term one needs isometries. More involved discussion of these issues is reserved for future work.

\section*{Acknowledgements}

\noindent This work was supported by the Russian state grant Goszadanie 3.9904.2017/8.9 and by the Foundation for the Advancement of Theoretical Physics and Mathematics ``BASIS'' and in part by the program of competitive growth of Kazan Federal University.

\providecommand{\href}[2]{#2}\begingroup\raggedright\endgroup

\end{document}